\documentclass[iop,apj,numberedappendix,appendixfloats,linenumbers]{emulateapj} 

\usepackage[export]{adjustbox}

\usepackage{amssymb}
\usepackage[usenames, dvipsnames]{color}

\usepackage{makecell}

\usepackage{color}
\usepackage{natbib}
\usepackage{verbatim}
\usepackage{times}
\usepackage{etoolbox}
\usepackage{amsmath}

\usepackage[citecolor=blue,hypertexnames=false]{hyperref}

\hypersetup{
    colorlinks=true,       
    linkcolor=blue,                 
    filecolor=magenta,      
    urlcolor=blue
}

\graphicspath {{Images/PDF_Files/}{Images/}}

\newcommand{\hi}{H~{\sc i}}
\newcommand{\hii}{H~{\sc ii}}
\newcommand{\htwo}{H$_2$}

\newcommand{\grays}{$\gamma$-rays}
\newcommand{\fermilat}{{\it Fermi}--LAT}

\newcommand{\p}{$\pm$ }

\submitted{Presented at the 8th International Fermi Symposium, Oct.~14-19, 2018, Baltimore, MD}

\shorttitle{\textit{Fermi}-LAT Observations Towards the Outer Halo of M31}
\shortauthors{Karwin, Murgia, Campbell, and Moskalenko}

\usepackage[pagewise]{lineno}

\begin{document}

\renewcommand{\linenumberfont}{\tiny}

\title{\textit{Fermi}-LAT Observations of $\gamma$-Ray Emission Towards the Outer Halo of M31}

\author{Christopher M. Karwin\altaffilmark{$\mathrm{\dagger}$}, Simona Murgia\altaffilmark{$\mathrm{\ddagger}$}, and Sheldon Campbell\altaffilmark{$\mathrm{\mathparagraph}$}}
\affil{\textit{Department of Physics and Astronomy, University of California, Irvine, CA 92697, USA}}

\author{Igor V. Moskalenko\altaffilmark{$\mathrm{*}$}}
\affil{\textit{Hansen Experimental Physics Laboratory and Kavli Institute for Particle Astrophysics and Cosmology, \\ Stanford University, Stanford, CA 94305, USA}}

\altaffiltext{$\mathrm{\dagger}$}{ckarwin@uci.edu}
\altaffiltext{$\mathrm{\ddagger}$}{smurgia@uci.edu}
\altaffiltext{$\mathrm{\mathparagraph}$}{sheldoc@uci.edu}
\altaffiltext{$\mathrm{*}$}{imos@stanford.edu}

\begin{abstract}
The Andromeda Galaxy is the closest spiral galaxy to us and has been the subject of numerous studies. It harbors a massive dark matter (DM) halo which may span up to $\sim$600 kpc across and comprises $\sim$90\% of the galaxy's total mass. This halo size translates into a large diameter of 42$^\circ$ on the sky for an M31--Milky Way (MW) distance of 785 kpc, but its presumably low surface brightness makes it challenging to detect with $\gamma$-ray telescopes. Using 7.6 years of \textit{Fermi} Large Area Telescope (\fermilat) observations, we make a detailed study of the $\gamma$-ray emission between 1--100 GeV towards M31's outer halo, with a total field radius of $60^\circ$ centered at M31, and perform an in-depth analysis of the systematic uncertainties related to the observations. We use the cosmic ray (CR) propagation code GALPROP to construct specialized interstellar emission models (IEMs) to characterize the foreground $\gamma$-ray emission from the MW, including a self-consistent determination of the isotropic component. We find evidence for an extended excess that appears to be distinct from the conventional MW foreground, having a total radial extension upwards of $\sim$120--200 kpc from the center of M31. We discuss plausible interpretations of the excess emission but emphasize that uncertainties in the MW foreground, and in particular, modeling of the \hi-related components, have not been fully explored and may impact the results. This study was first presented in a poster at the 8th International Fermi Symposium. The article describing the full analysis is under preparation and will be published elsewhere.

\end{abstract}

\section{Introduction}

\setcounter{footnote}{0}

\begin{figure*}
\centering
\includegraphics[width=0.49\textwidth]{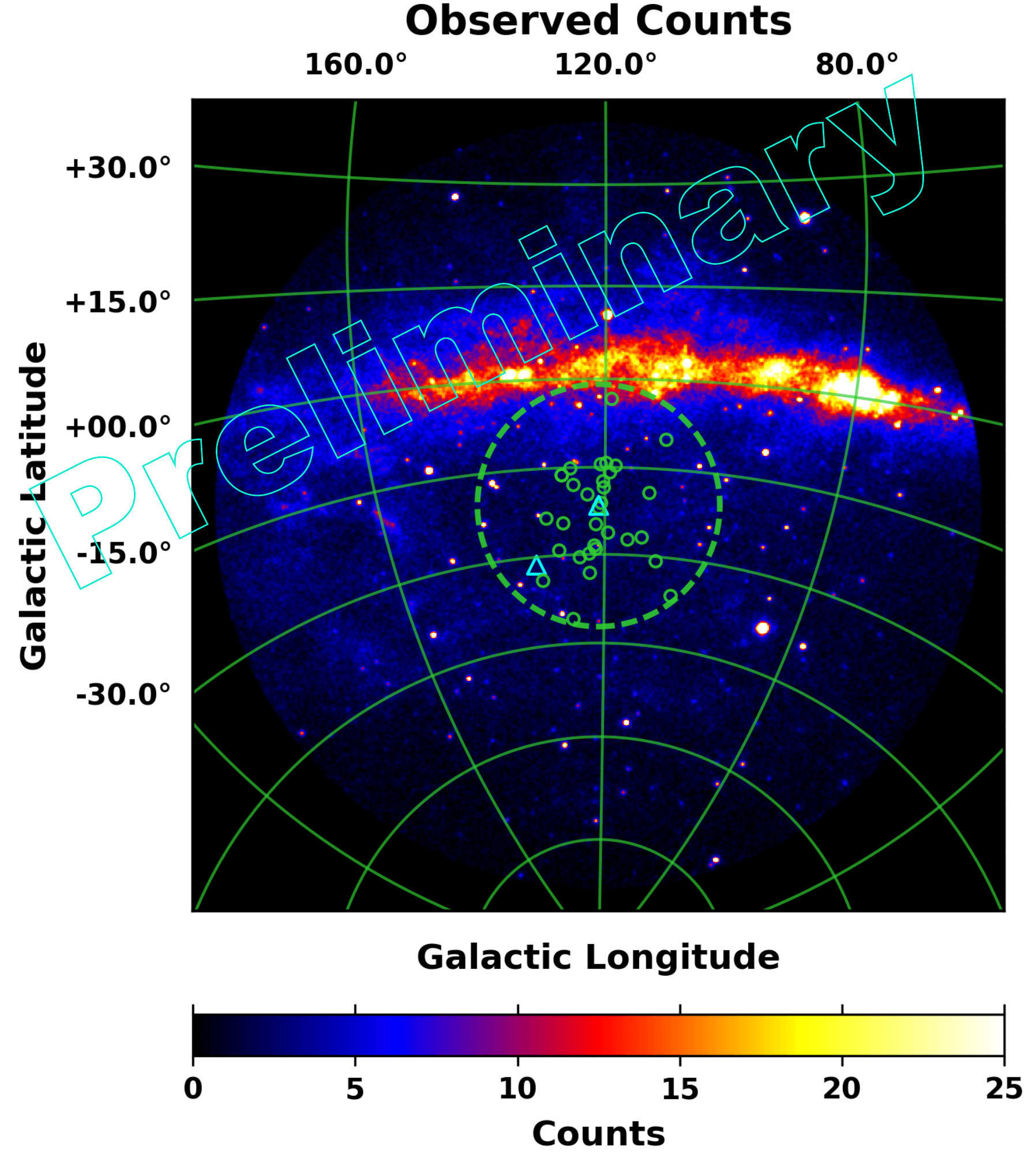}
\includegraphics[width=0.49\textwidth]{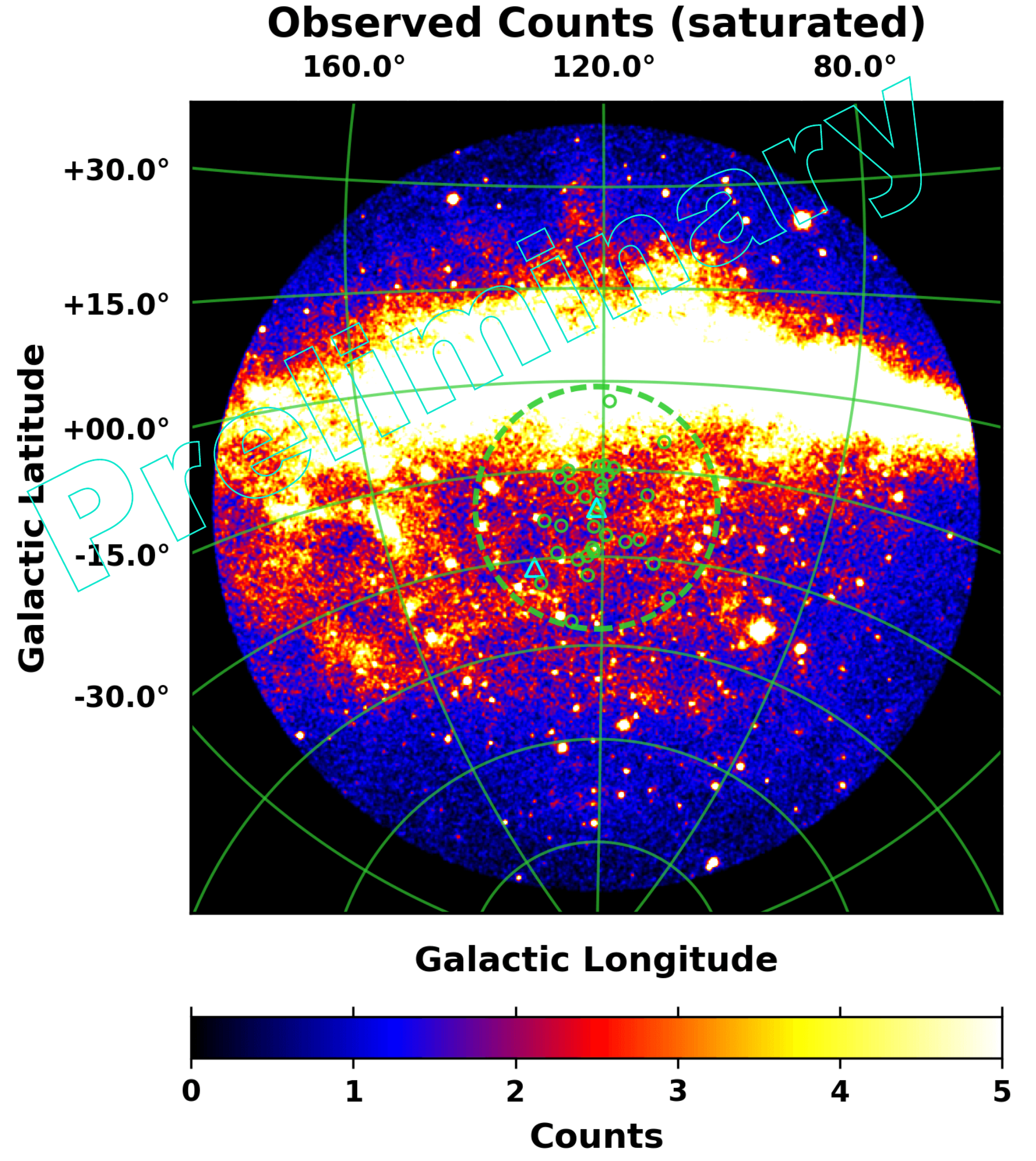}
\caption{Observed counts (left) and saturated counts (right) for a $60^\circ$ radius centered at M31, and an energy range of 1--100 GeV. The green dashed circle ($21^\circ$ in radius) corresponds to a 300 kpc projected radius centered at M31, for an M31--MW distance of 785 kpc, i.e. the canonical virial radius of M31. Also shown is M31's population of dwarf galaxies. M31 and M33 are shown with cyan triangles, and the other dwarfs are shown with $1^\circ$ green circles, each centered at the optical center of the respective galaxy. The sizes of the circles are a bit arbitrary, although they roughly correspond to the point spread function (PSF) of \textit{Fermi}-LAT, which at 1 GeV is $\sim$$1^\circ$. Most of the MW dwarfs are not detected by \textit{Fermi}-LAT, and so we do not necessarily expect the individual M31 dwarfs to be detected. The primary purpose of the overlay is to provide a qualitative representation of the extent of M31's outer halo, and to show its relationship to the MW disk. Note that $\sim$3 dwarfs (which are thought to be gravitationally bound to M31) have been observed as far as $\sim$300 kpc, as can be seen in the figure.} 
\label{fig:observed_counts}
\end{figure*}
\begin{figure}
\centering

\includegraphics[width=0.47\textwidth]{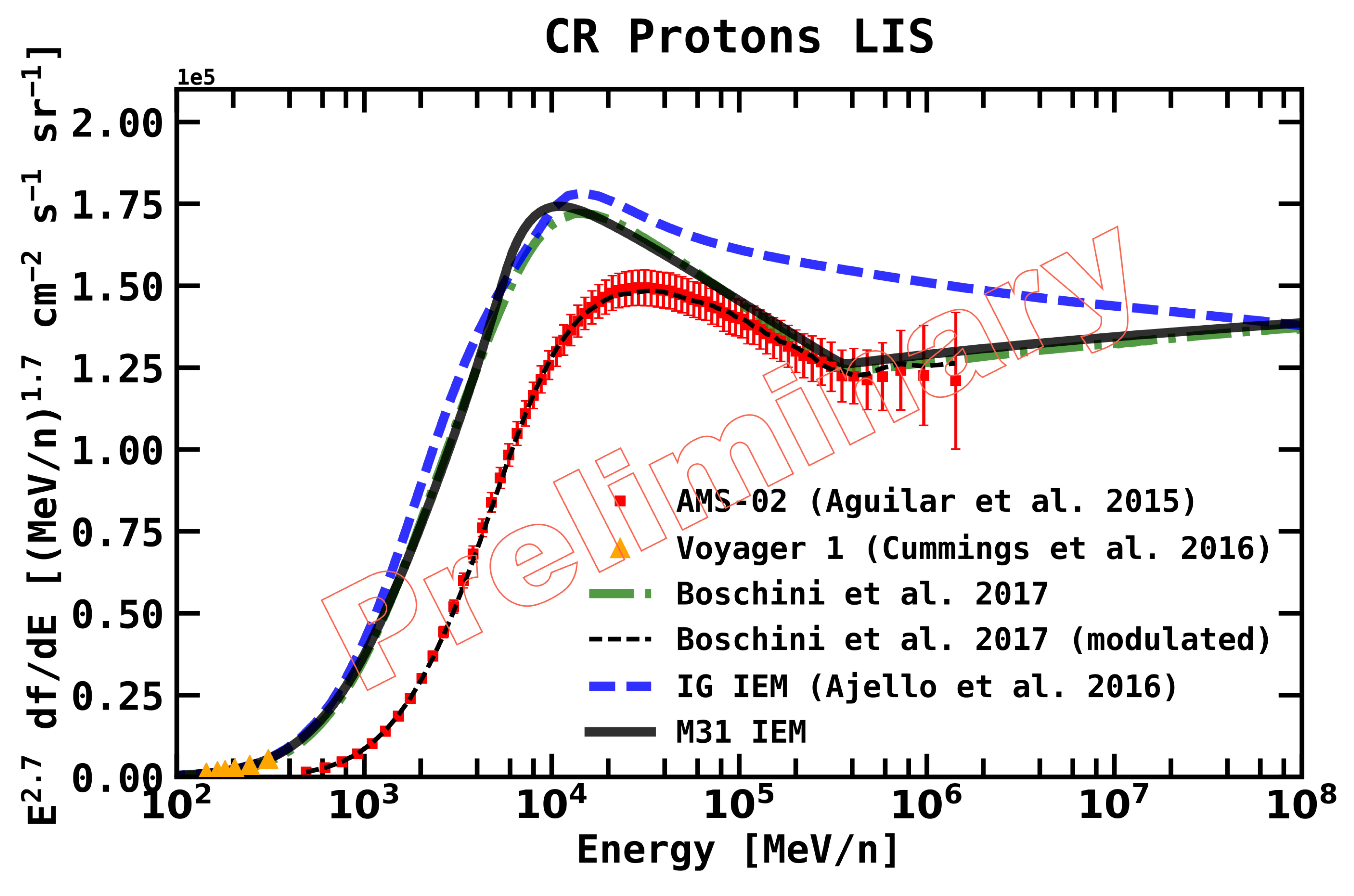}
\includegraphics[width=0.47\textwidth]{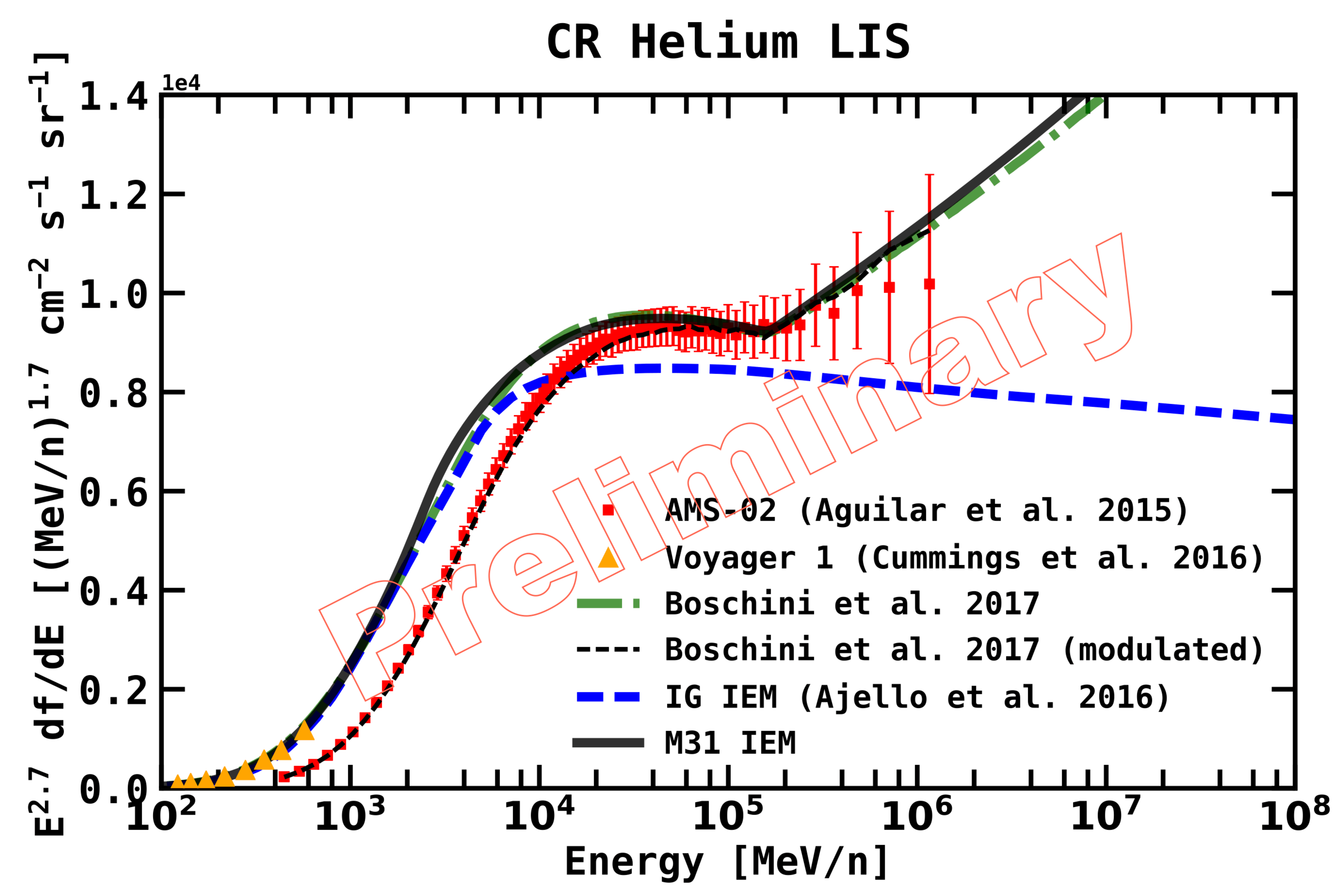}
\includegraphics[width=0.47\textwidth]{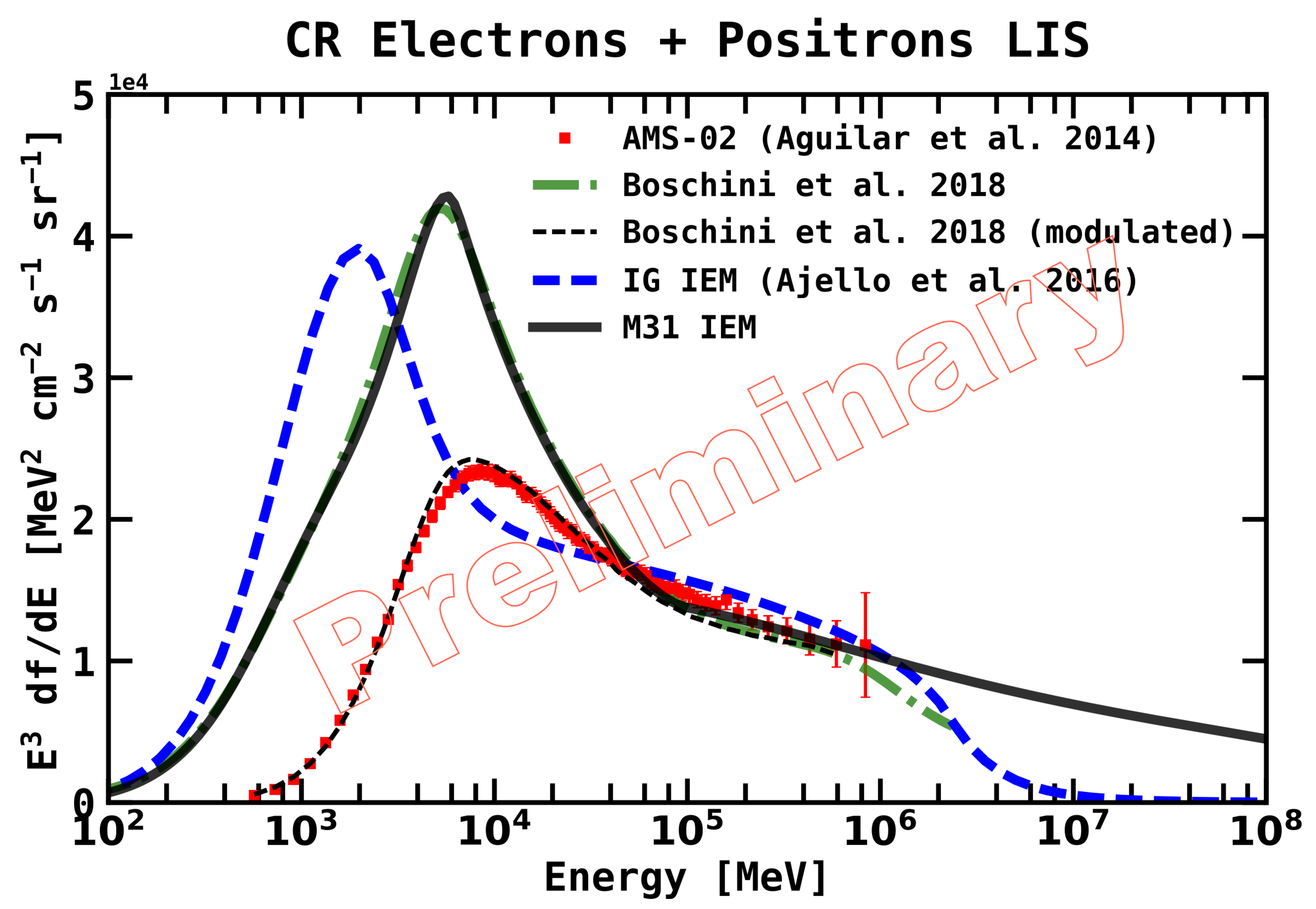}

\caption{The LIS for CR protons (top), He (middle), and all electrons ($e^- +e^+$) (bottom). The latest AMS-02 measurements from~\citet{PhysRevLett.113.221102,PhysRevLett.114.171103,PhysRevLett.115.211101} are shown with red squares. The green dashed line shows the results from~\citet{Boschini:2017fxq,Boschini:2018zdv}, which employ GALPROP and HelMod together in an iterative manner to derive the LIS. We adopt their derived GALPROP CR parameters, and the LIS for our IEM (M31 IEM: solid black line) are roughly the same. The thin dotted black line shows the LIS modulated with HelMod \citep{Boschini:2017fxq,Boschini:2018zdv}. Yellow triangles show the Voyager 1 proton and He data in the local interstellar medium \citep{2016ApJ...831...18C}. Voyager 1 electron data are below 100 MeV and, therefore, are not shown. In addition we show the LIS for the (``Yusifov'') IEM in~\citet{TheFermi-LAT:2015kwa}, which we use as a reference model in our study of the systematics for the M31 field.}
\label{fig:CR_LIS}
\end{figure}

The goal of this work is to search for extended $\gamma$-ray emission originating beyond the galactic disk of M31, and to examine the implications for CRs and DM. There are two primary motivations for this search. First, CR interactions with the radiation field of M31's stellar halo and/or the circumgalactic gas could generate a detectable signal in $\gamma$-rays. Secondly, M31's DM halo has a large extension on the sky and could produce a detectable signal within currently allowed DM scenarios, which would be complementary to other targets, and specifically, the Galactic center. What makes M31 advantages in regards to DM searches with $\gamma$-rays is that the entire DM halo is seen from the outside, so we see the extended integral signal. For the MW we see through the halo, and so it can be easily confused with diffuse components. Our primary field of interest is a $28^\circ \times 28^\circ$ square region, which amounts to a projected radius of $\sim$200 kpc from the center of M31. Our study complements previously published results on M31~\citep{Fermi-LAT:2010kib,ogelman2010discovery,Pshirkov:2015hda,Pshirkov:2016qhu,Ackermann:2017nya} and is the first to explore the farthest reaches of the M31 system in $\gamma$-rays.

\section{Observations}

Our full region of interest (ROI) corresponds to a radius of $60^\circ$ centered at the position of M31, $(l,b) = (121.17^{\circ}, -21.57^{\circ})$. We employ front and back converting events corresponding to the P8R2\_CLEAN\_V6 selection. The events have energies in the range 1--100 GeV and have been collected from 2008-08-04 to 2016-03-16 (7.6 years). The data are divided into 20 bins equally spaced in logarithmic energy, with $0.2^\circ\times0.2^\circ$ pixel size. The analysis is carried out with the \fermilat\ ScienceTools  (version v10r0p5)\footnote{Available at \url{http://fermi.gsfc.nasa.gov/ssc/data/analysis}}. In  particular, the binned maximum likelihood fits are performed with the {\it gtlike} package.

Figure~\ref{fig:observed_counts} shows the total observed counts between 1--100 GeV for the full ROI. Two different count ranges are displayed. The map on the left shows the full range. The bright emission along 0$^\circ$ latitude corresponds to the plane of the MW. The map on the right shows the saturated counts map, emphasizing the lower counts at higher latitudes.  Overlaid is a green dashed circle ($21^\circ$ in radius) corresponding to a 300 kpc projected radius centered at M31, for an M31-MW distance of 785 kpc, i.e.~the canonical virial radius of M31. Also shown is M31's population of dwarf galaxies. The primary purpose of the overlay is to provide a qualitative representation of the extent of M31's outer halo, and to show its relationship to the MW disk. Note that we divide the full ROI into subregions, and our primary field of interest is a $28^\circ \times 28^\circ$ square region centered at M31, which we refer to as field M31 (FM31).

\section{Building the Interstellar Emission Models}

The foreground emission from the MW and the isotropic component (the latter includes unresolved extragalactic diffuse $\gamma$-ray emission, residual instrumental background, and possibly contributions from other Galactic components which have a roughly isotropic distribution) are the dominant  contributions in \grays\ towards the M31 region. We use the CR propagation code GALPROP\footnote{Available at \url{https://galprop.stanford.edu}}(v56) \citep{Moskalenko:1997gh,Moskalenko:1998gw,Strong:1998pw,Strong:1998fr,2006ApJ...642..902P,Strong:2007nh,Vladimirov:2010aq,Johannesson:2016rlh,porter2017high,Johannesson:2018bit,PhysRevC.98.034611} to construct specialized interstellar emission models (IEMs) to characterize the MW foreground emission, including a self-consistent determination of the isotropic component. These foreground models are physically motivated and \emph{are not} subject to the same caveats\footnote{The list of caveats on the \fermilat\ diffuse model is available at \url{https://fermi.gsfc.nasa.gov/ssc/data/analysis/LAT_caveats.html} \label{caveats}} for extended source analysis as the FSSC IEM provided by the \fermilat\ Collaboration for point source analysis~\citep{Acero:2016qlg}.

\begin{figure*}
\centering
\includegraphics[width=0.98\textwidth]{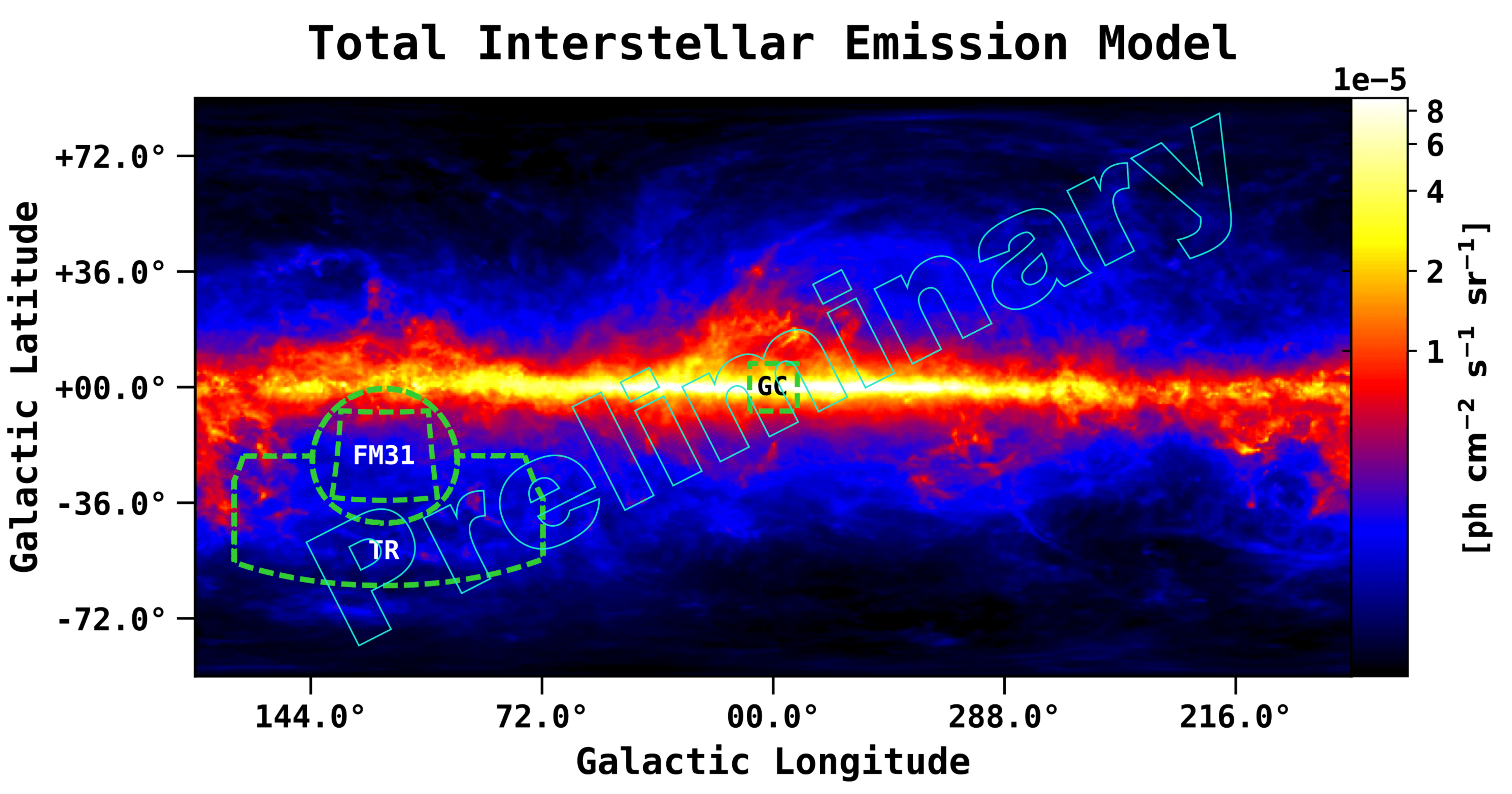}
\caption{The total interstellar emission model (IEM) for the MW integrated in the energy range 1--100 GeV. The color corresponds to the intensity, and is shown in logarithmic scale. The intensity level is for the initial GALPROP output, before tuning to the $\gamma$-ray data. The map is shown in a Plate Carr\'{e}e projection, and the pixel size is 0.25 deg/pix. The model has contributions from $\pi^0$-decay, (anisotropic) IC emission, and Bremsstrahlung. Overlaid is the region of interest (ROI) used in this analysis. From the observed counts (Figure~\ref{fig:observed_counts}) we cut an $84^\circ\times84^\circ$ ROI, which is centered at M31. The green dashed circle is the 300 kpc boundary corresponding to M31's canonical virial radius (of $\sim$$21^\circ$), as also shown in Figure~\ref{fig:observed_counts}. We label the field within the virial radius as field M31 (FM31), and the region outside (and south of latitudes of $-21.57^\circ$) we label as the tuning region (TR). Longitude cuts are made on the ROI at $l=168^\circ \ \mathrm{and} \ l=72^\circ$. For reference we also show the Galactic center region (GC), which corresponds to a $15^\circ\times15^\circ$ square centered at the GC.}
\label{fig:galactic_diffuse_schematic}
\end{figure*}

\begin{figure}
\centering
\includegraphics[width=0.48\textwidth]{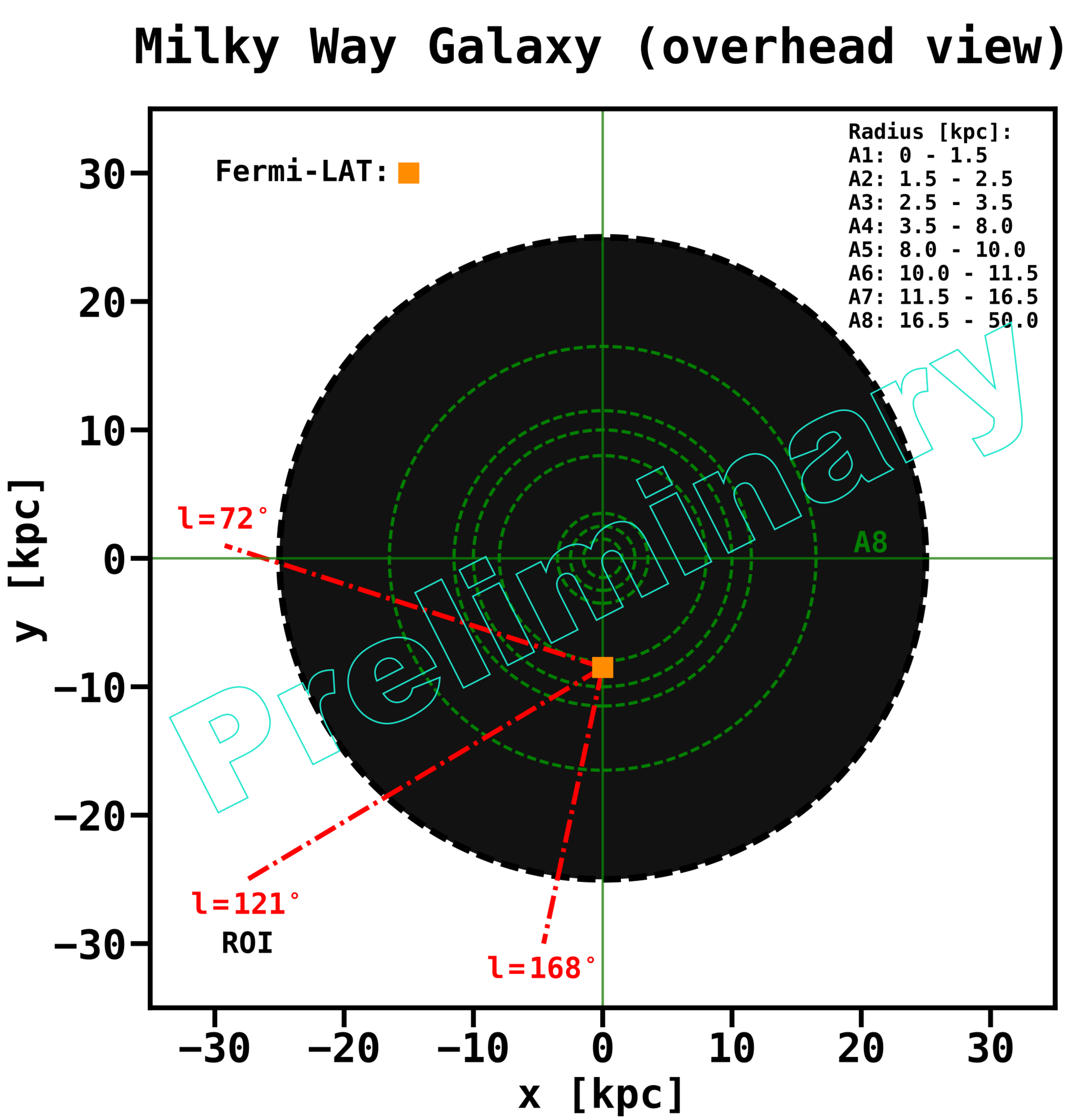}
\caption{Schematic of the eight concentric circles which define  the annuli (A1--A8) in the IEM. The  ranges in Galactocentric radii are reported in the legend. Note that the full extension of A8 is not shown. Only A5--A8 contribute to the Galactic foreground emission for the field used in this analysis.}
\label{fig:measurement_schematic}
\end{figure}

The parameters of the GALPROP model are tuned to the measured local interstellar spectra (LIS) of CRs, including the latest AMS-02 measurements. We have adopted the best-fit parameters from the tuning procedure performed in~\citet{Boschini:2017fxq,Boschini:2018zdv}, where GALPROP and HelMod\footnote{Available at \url{http://www.helmod.org/}} are implemented in an iterative manner, thereby accounting for solar modulation in a physically motivated way when fitting to the local CR measurements. This is summarized in Figure~\ref{fig:CR_LIS}.

Figure~\ref{fig:galactic_diffuse_schematic} shows the total interstellar emission model, which consists of individual components for $\pi^0$-decay, inverse Compton (IC), and Bremsstrahlung. The components are defined in Galctocentric annuli, as depicted in Figure~\ref{fig:measurement_schematic}. In total there are 8 annuli, but for FM31 only annulus 5 (the local annulus) and beyond contribute to the foreground emission. FM31 has a significant emission associated with \hi\ gas, but there is very little emission from \htwo{} gas. A uniform spin temperature of 150 K is assumed for the baseline IEM. The foreground emission from \hii\ and Bremsstrahlung are subdominant. Our model also accounts for the dark neutral medium (DNM). The anisotropic formalism is employed for the calculation of the IC component. In addition to the GALPROP IEM components, we also use a template approach to account for inaccuracies in the foreground model relating to the neutral gas along the line of sight. To model the point sources in the region, we employ the 3FGL as a starting point, and because of the larger statistics of our data set, we account for additional point sources self-consistently with the M31 IEM by implementing a point source finding procedure, which is based on a wavelet transform algorithm.

\begin{figure}
\centering
\includegraphics[width=0.49\textwidth]{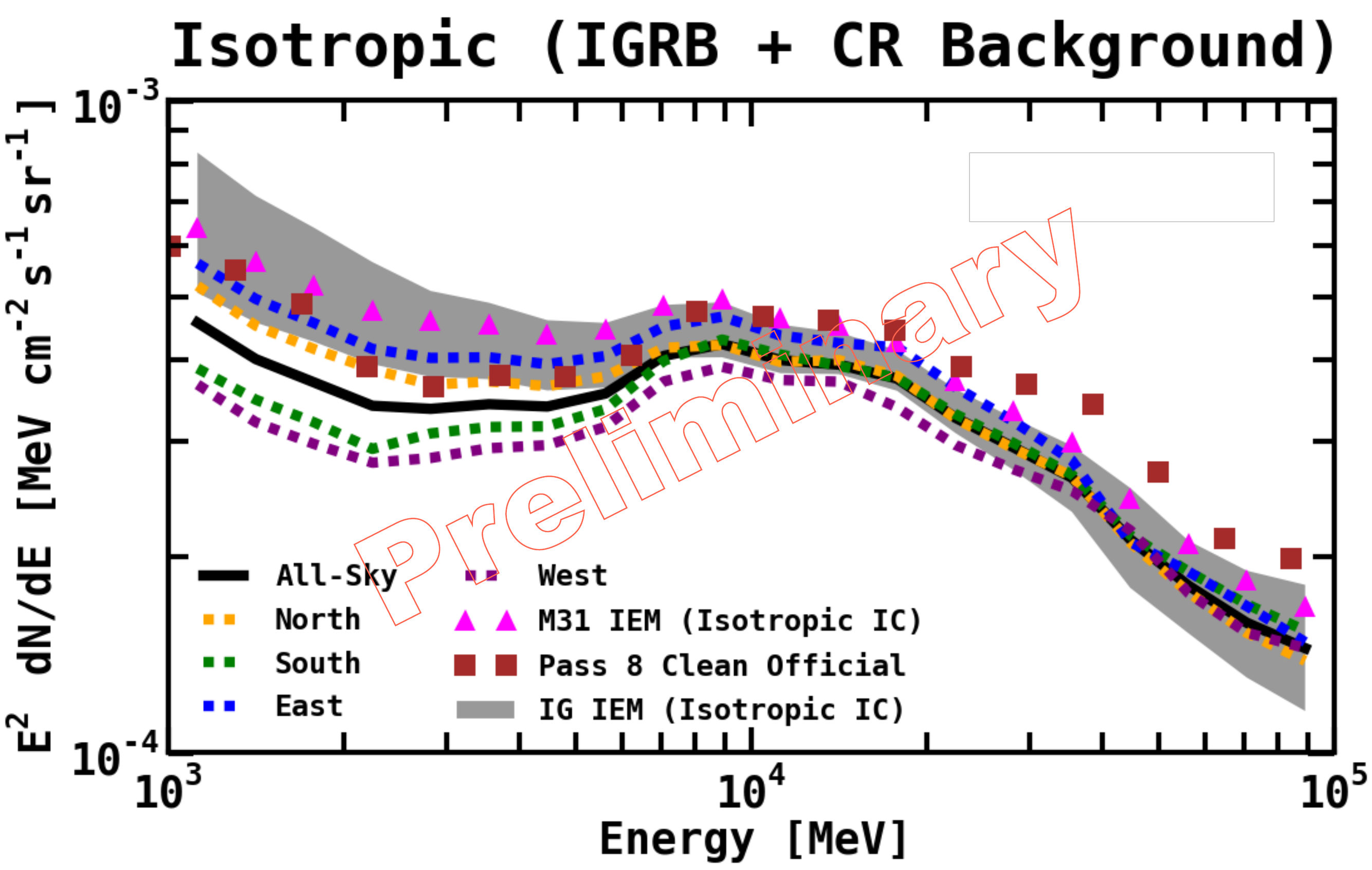}
\caption{The spectrum of the isotropic component has a dependence on the IEM and the ROI used for the calculation, as well as the data set. For the M31 IEM (which uses the anisotropic IC sky maps) we calculate the \textbf{All-Sky} (solid black line) isotropic component in the following region: $|b| \geq 30^\circ, \ 45^\circ \leq  l \leq 315^\circ$. We also calculate the isotropic component in the different sky regions: \textbf{north}: $b\geq 30^\circ, \ 45^\circ \leq  l \leq 315^\circ$ (orange dashed line); \textbf{south}:  $b\leq -30^\circ,\ 45^\circ \leq  l \leq 315^\circ$ (green dashed line); \textbf{east}: $|b|\geq 30^\circ, \ 180^\circ \leq  l \leq 315^\circ$ (blue dashed line); and \textbf{west}: $|b|\geq 30^\circ, \ 45^\circ \leq  l \leq 180^\circ$ (purple dashed line). Magenta triangles show the all-sky isotropic component for the M31 IEM derived using the isotropic IC formalism. The brown squares show the official FSSC isotropic spectrum (iso\_P8R2\_CLEAN\_V6\_v06). The grey band is our calculated isotropic systematic uncertainty for the inner Galaxy IEM (which uses the isotropic IC formalism).}
\label{fig:Isotropic_Sytematics}
\end{figure}

\begin{figure*}
\centering
\includegraphics[width=0.48\textwidth]{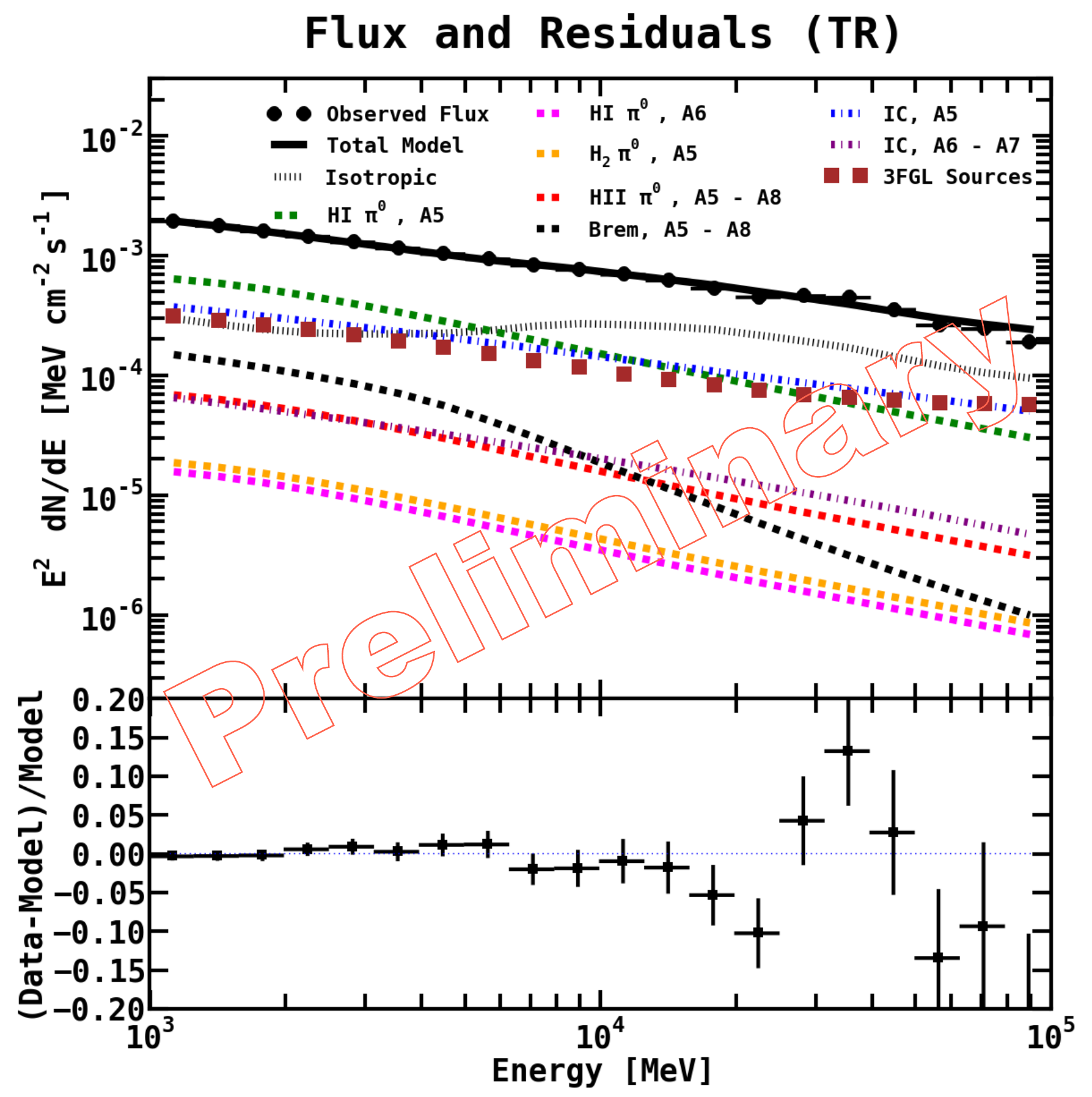}
\includegraphics[width=0.48\textwidth]{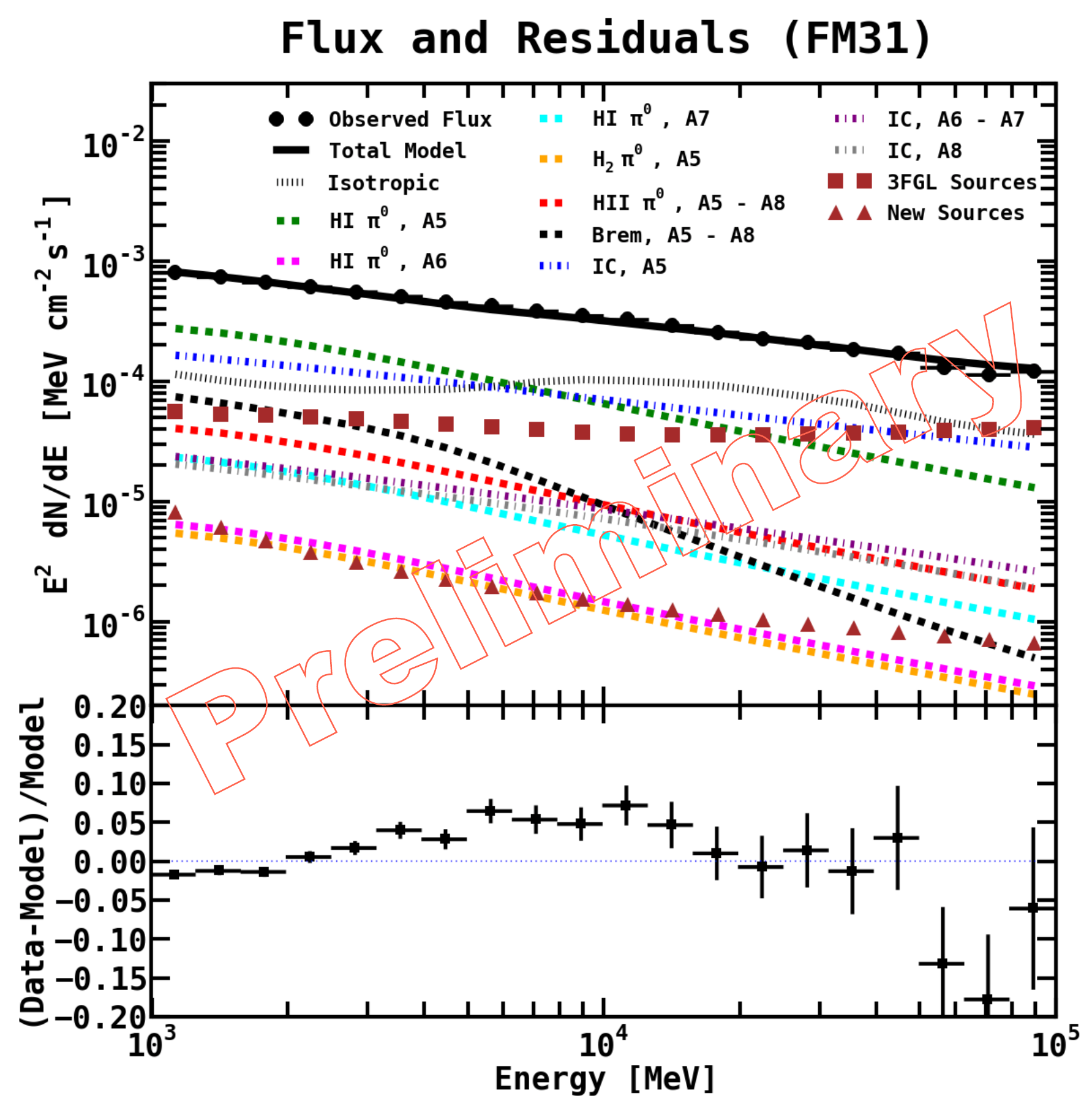}
\caption{\textbf{Left:} Flux (upper panel) and fractional count residuals (lower panel) for the fit in the TR. The components of the IEM are listed in the legend. The residuals show fairly good agreement over the entire energy range. \textbf{Right:} Flux (upper panel) and fractional count residuals (lower panel) for the baseline fit in FM31. The fractional residuals show an excess between $\sim$3--20 GeV reaching a level of $\sim$5\%. Above and below this range the data is being over-modeled as the fit tries to balance the excess with the negative residuals.}
\label{fig:flux_and_residuals_TR}
\end{figure*}

\begin{figure*}
\centering
\includegraphics[width=0.33\textwidth]{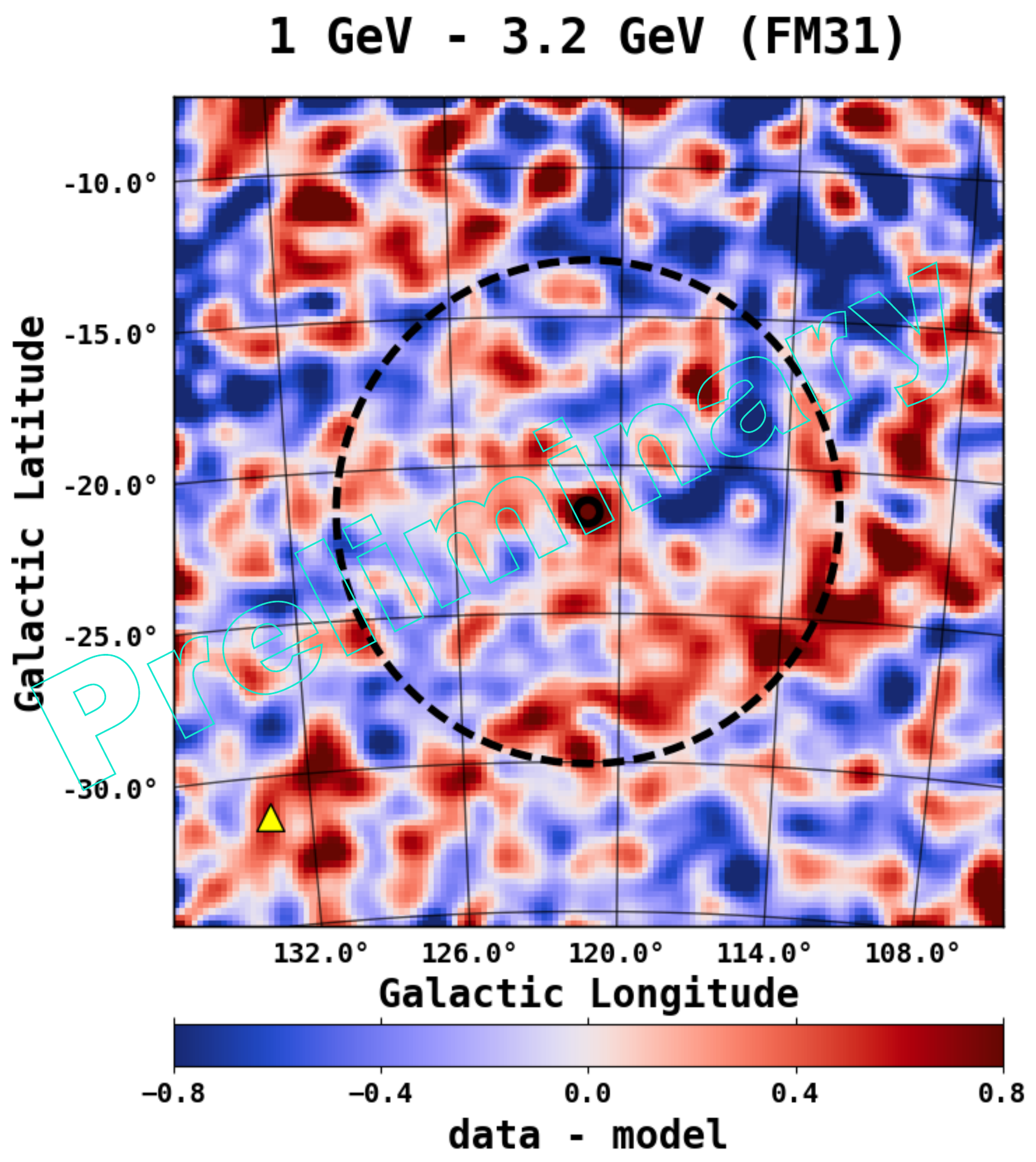}
\includegraphics[width=0.33\textwidth]{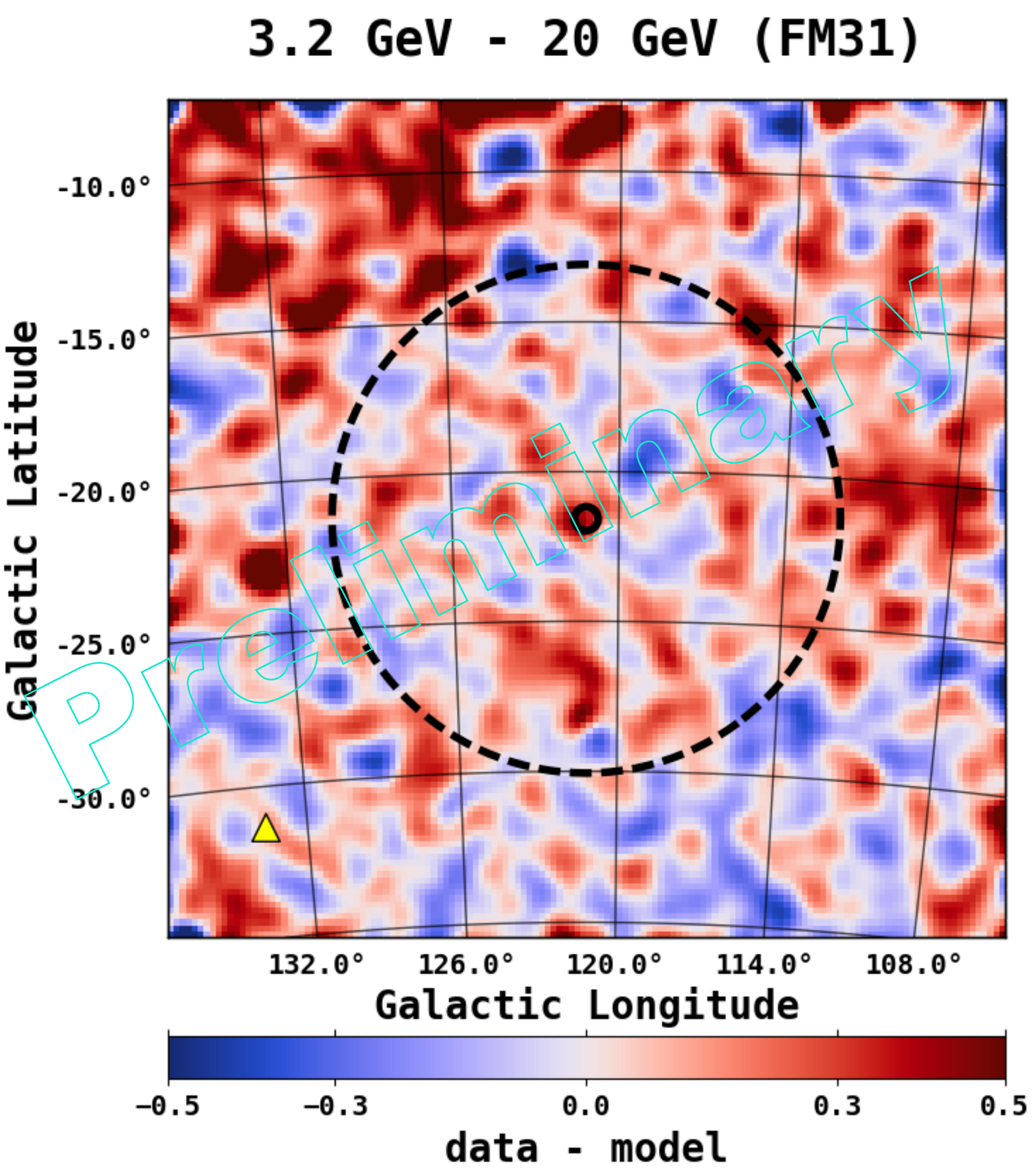}
\includegraphics[width=0.33\textwidth]{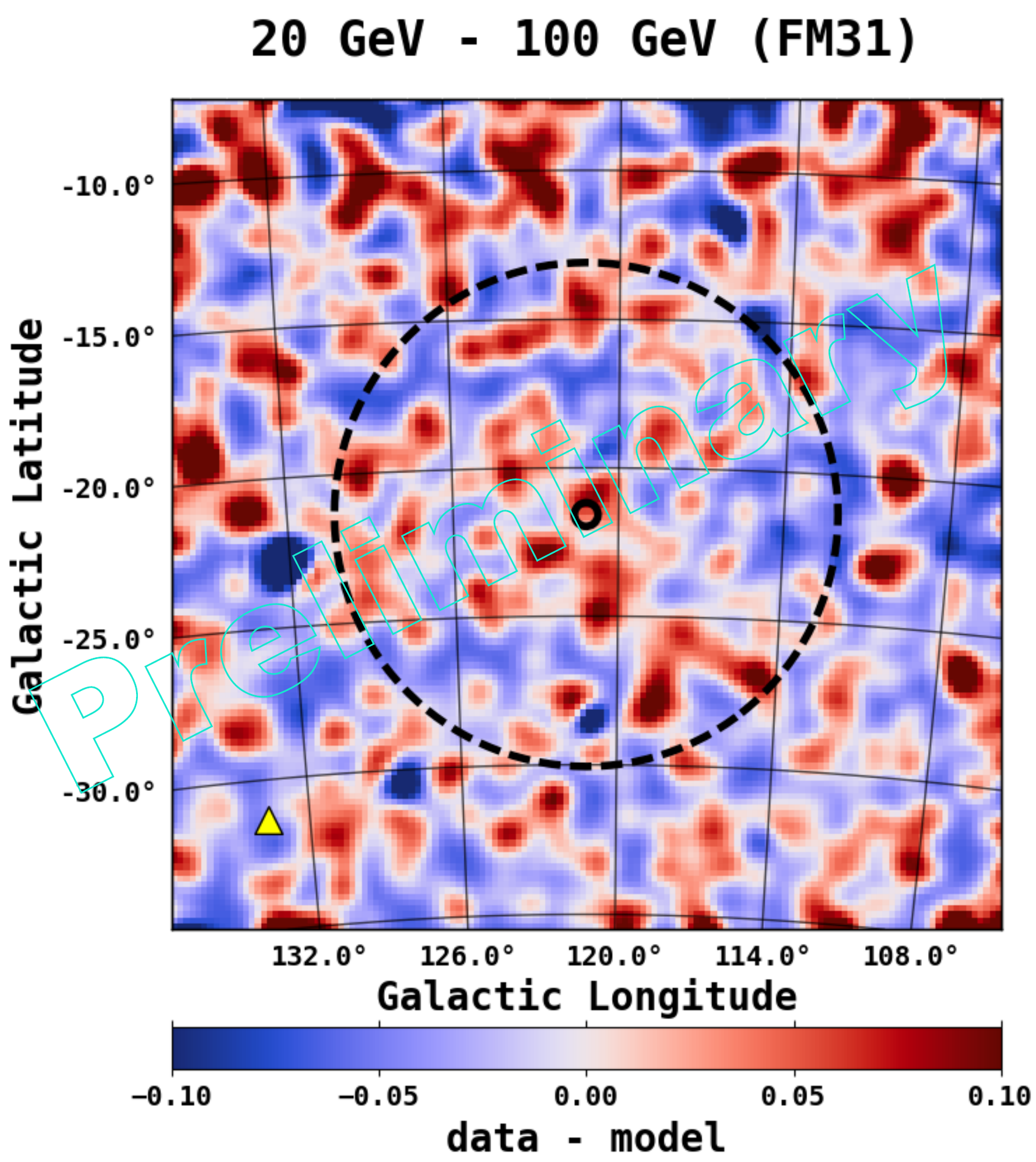}
\caption{Spatial count residuals (data $-$ model) resulting from the baseline fit in FM31 for three different energy bands, as indicated above each plot. The energy bins are chosen to coincide with the excess observed in the fractional residuals. The color scale corresponds to counts/pixel, and the pixel size is $0.2^\circ \times 0.2^\circ$. The images are smoothed using a $1^\circ$ Gaussian kernel. This value corresponds to the PSF (68\% containment angle) of \textit{Fermi}-LAT, which at 1 GeV is $\sim$$1^\circ$. For reference, the position of M33, $(l,b) = (133.61^\circ, -31.33^\circ)$, is shown with a yellow triangle. We eventually add to the model three symmetric M31-related templates, as discussed in the text. The boundaries for these templates are overlaid to the residual maps. The solid black circle ($0.4^\circ$ in radius) shows the boundary for the inner galaxy template. The dashed black circle ($8.5^\circ$ in radius) shows the boundary for the spherical halo template, which corresponds to a projected radius of 117 kpc. The far outer halo template covers the remaining extent of the field. Further details are given in Figure~\ref{fig:M31_radial_profile}.}
\label{fig:spatial_residuals_FM31_tuned}
\end{figure*}

\begin{figure*}
\centering
\includegraphics[width=0.44\textwidth]{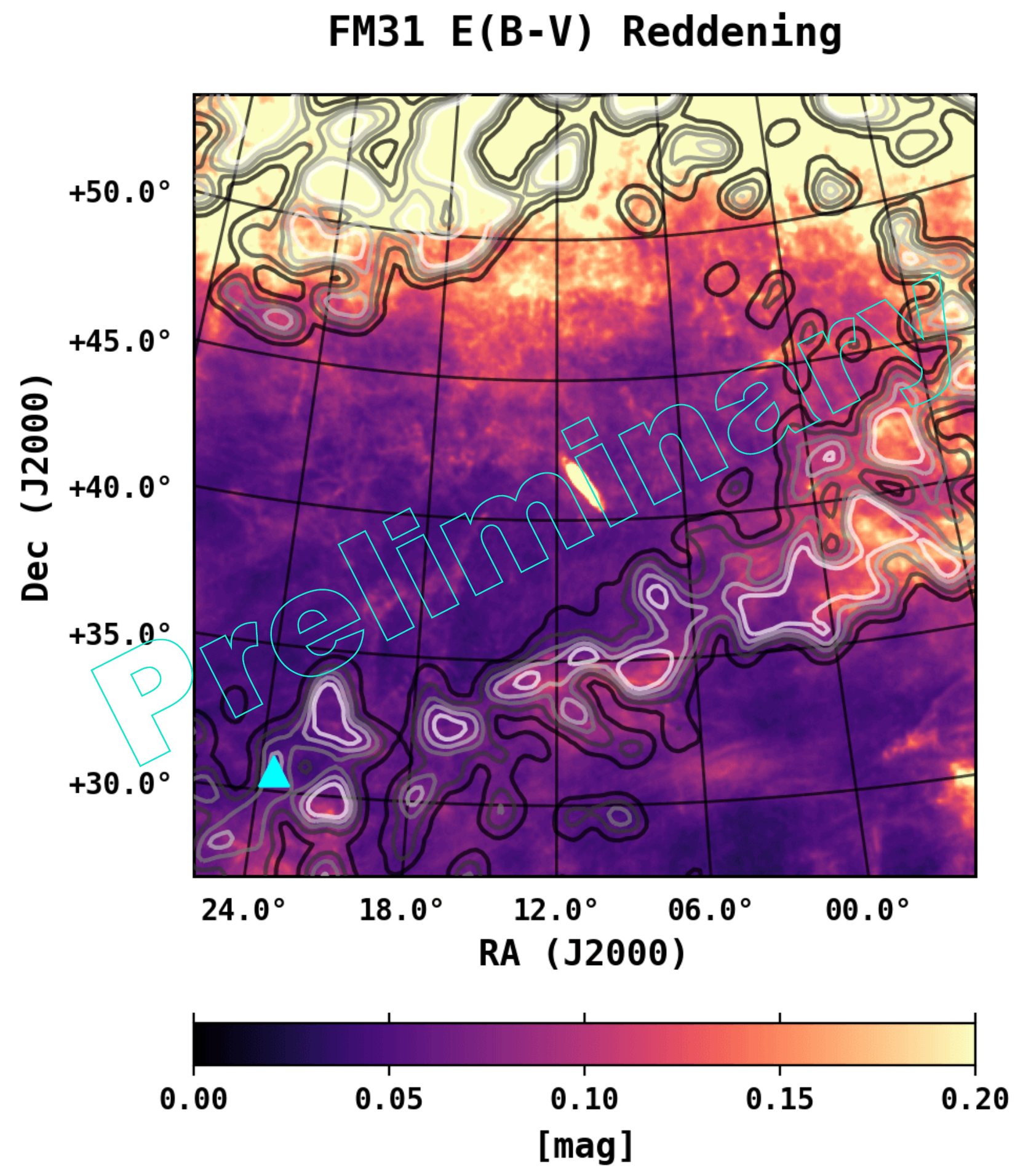}
\includegraphics[width=0.49\textwidth]{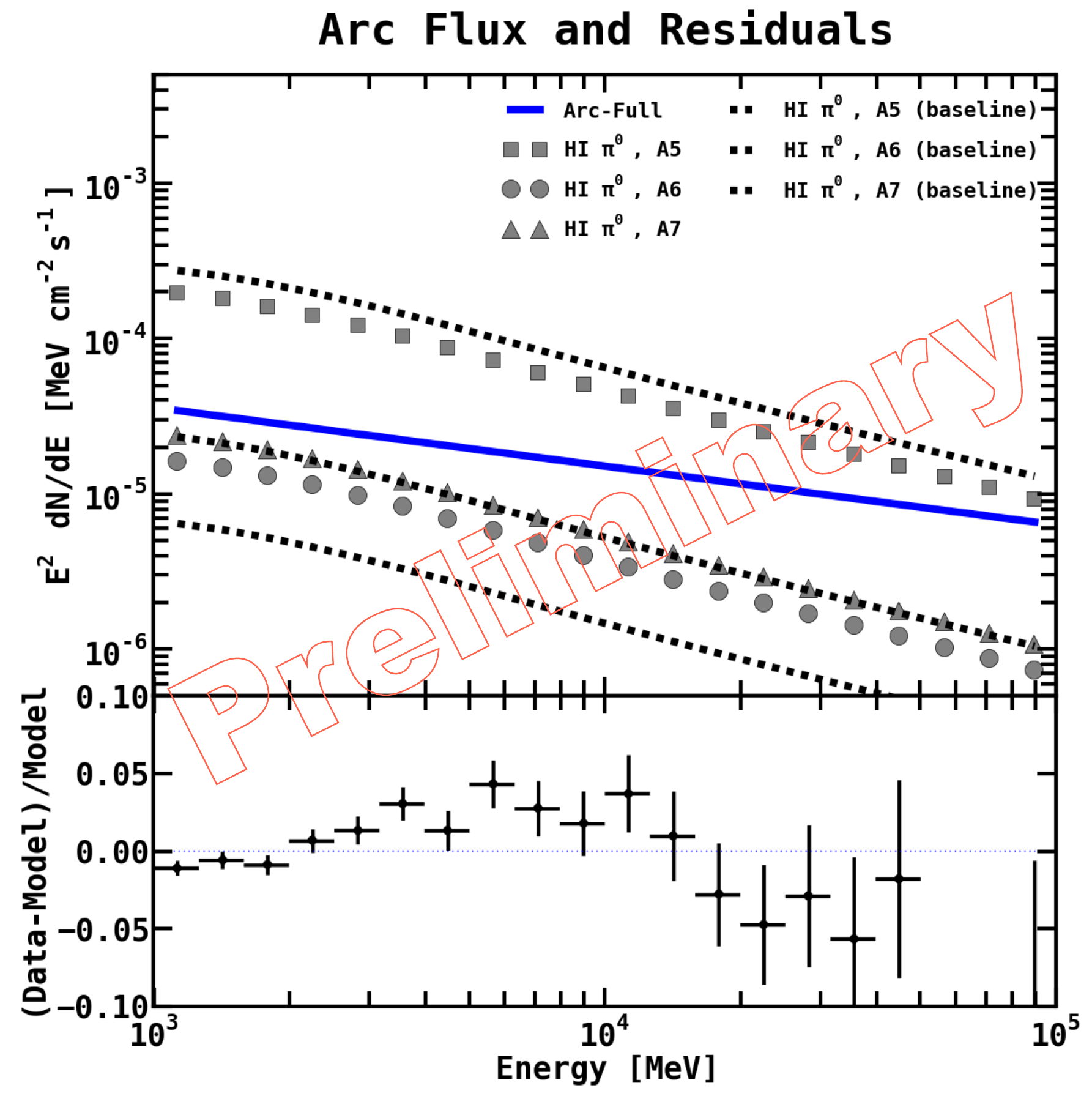}
\caption{\textbf{Left:} The dust reddening map for FM31 from~\citet{schlegel1998maps}. Overlaid are contours for the arc template, which is added to the model to account for the arc feature observed in the residuals, likely related to inaccuracies in the foreground model (at least the upper portion of the arc). The cyan triangle shows the (projected) position of M33. \textbf{Right:} Spectra and fractional energy residuals resulting from the arc fit. The arc component is given a power law spectrum, and the spectral parameters are fit simultaneously with the other components in the region, just as for the baseline fit. The blue solid line is the best-fit spectrum for the arc template. The bottom panel shows the remaining fractional residuals. The arc template is unable to flatten the excess between $\sim$3--20 GeV.}
\label{fig:FM31_dust}
\end{figure*}

We calculate the isotropic component self-consistently with the M31 IEM. The main calculation is performed over the full sky in the following region: $|b| \geq 30^\circ, \ 45^\circ \leq  l \leq 315^\circ$. In addition, we also calculate the isotropic spectrum in different subregions. The results are summarized in Figure~\ref{fig:Isotropic_Sytematics}. To better determine the normalization of the isotropic component we use a tuning region (TR) directly below FM31, outside of the virial radius. The best-fit normalization is found to be 1.06 \p 0.04, and this remains fixed for all other fits with the M31 IEM. The fit in the TR yields a model that describes the data well across the entire region and at all energies. The best-fit normalizations of the IEM components in the TR are all in reasonable agreement with the GALPROP predictions. The spectra for the IEM components and the remaining fractional energy residuals for the fit in the TR are shown in the left panel of Figure~\ref{fig:flux_and_residuals_TR}.

\section{Results for the M31 Field}

\begin{figure*}
\centering
\includegraphics[width=0.49\textwidth]{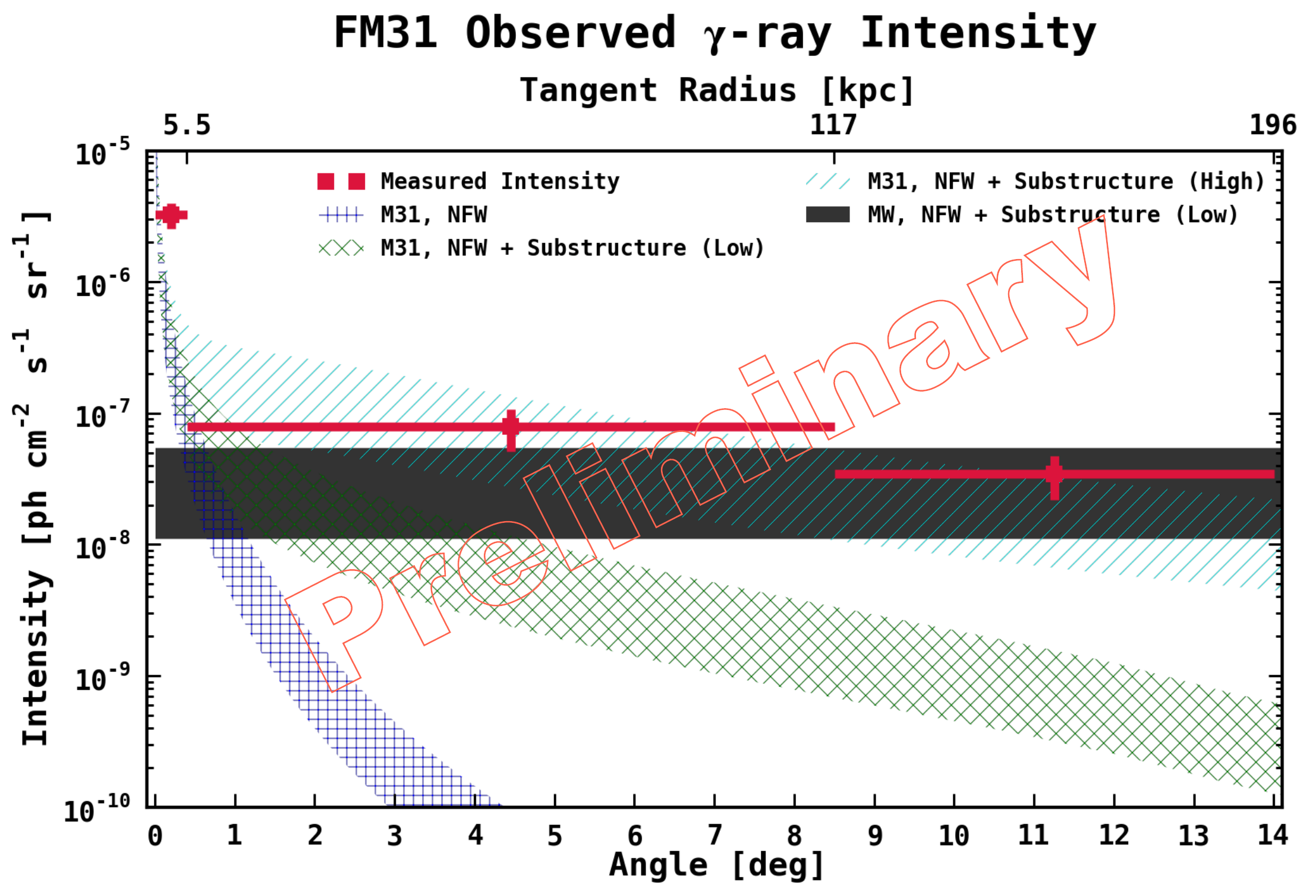}
\includegraphics[width=0.47\textwidth]{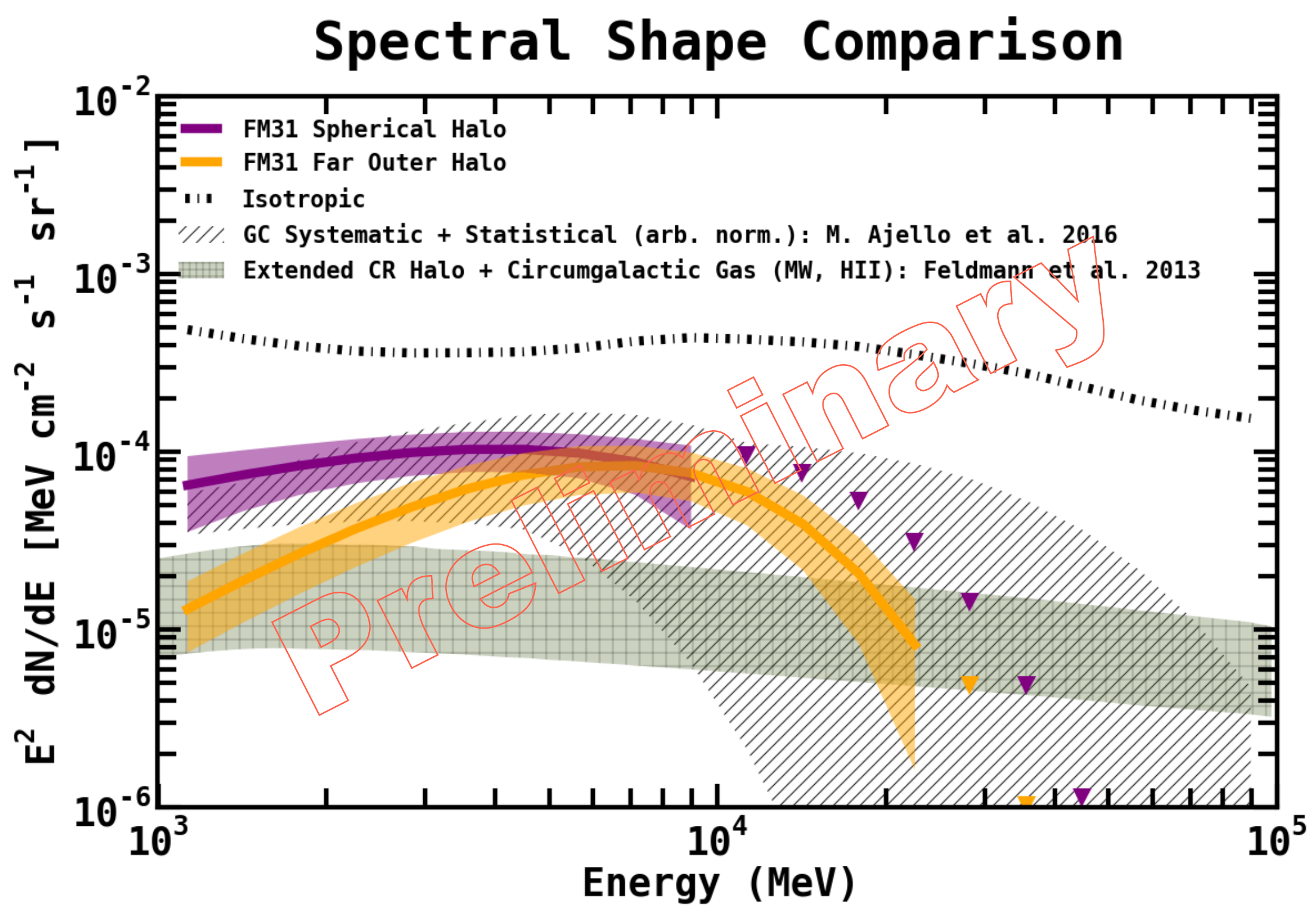}
\caption{We model M31 and its outer halo with three symmetric uniform templates centered at M31. The inner disk (inner galaxy) has a radial extension of $0.4^\circ$ (5.5 kpc projected radius). The intermediate ring (spherical halo) has a radial extension from $0.4^\circ < r \leq 8.5^\circ$ (117 kpc projected radius), and it encloses a majority of M31's globular cluster population and dwarf galaxy population, as well as a large lopsided H I cloud centered in projection on M31, possibly associated with the M31 system, i.e. the M31 cloud. The outer ring (far outer halo) covers the remaining extent of our primary M31 field (FM31), corresponding to a total projected radius of $\sim$200 kpc. The boundaries for the M31-related components are shown in Figure~\ref{fig:spatial_residuals_FM31_tuned}. The fit also includes the arc template. The M31-related components (inner galaxy, spherical halo, and far outer halo) are detected at the significance levels of 7$\sigma$, 7$\sigma$, and 5$\sigma$, respectively. We discuss plausible interpretations of the observed residual emission, but emphasize that uncertainties in the MW foreground, and in particular modeling of the H I-related emission, have not been fully explored and may impact the results. \textbf{Left:} Radial intensity profile for the M31-related components. For reference, we compare the radial profile to expectations for DM annihilation along the line of sight, including the contributions from both the M31 halo and the MW halo. \textbf{Right:} Spectral shape comparison to the Galactic center excess (for an arbitrary normalization), as observed in~\citet{TheFermi-LAT:2015kwa}. Also shown is a prediction for CRs interacting with the ionized gas of the circumgalactic medium from~\citet{Feldmann:2012rx}. Note that the prediction is for a MW component, but we are primarily interested in a spectral shape comparison. We consider this because it is a possible source of $\gamma$-ray emission in the region, but based on the properties of the observed excess, we find it seemingly unlikely that the corresponding emission is dominated by these types of interactions. For reference, the isotropic component is plotted as well.}
\label{fig:M31_radial_profile}
\end{figure*}

The baseline fit in FM31 results in positive residual emission in the fractional energy residuals between $\sim$3--20 GeV at the level of $\sim$5\%, as seen in the right panel of Figure~\ref{fig:flux_and_residuals_TR}. Figure~\ref{fig:spatial_residuals_FM31_tuned} shows the corresponding spatial residuals for three different energy bands. The bands are chosen to coincide with the positive residual emission observed in the fractional  energy residuals. The spatial residuals show structured excesses and deficits, primarily at lower energies ($\sim$1--3 GeV). In the second energy bin some of these structures can also be seen, but overall the residual emission is more uniformly distributed than for the first energy bin. 

A significant fraction of the structured excess emission in FM31 is found to be spatially correlated with the \hi\ column density and the foreground dust, including regions where the dust is relatively cold. This may be indicative of a spatially varying spin temperature. Correspondingly, the structured residual emission may be related to inaccuracies in the modeling of the DNM, which is determined as part of an all-sky procedure. We, therefore, refine the baseline IEM by constructing a template to account for potential mis-modeling of these components. The template is obtained by selecting the excess emission in FM31 that correlates with \hi\ tracers. We refer to this as the arc template, and the left panel of Figure~\ref{fig:FM31_dust} shows the corresponding contours overlaid to the dust reddening map from~\citet{schlegel1998maps}. This procedure accounts for any un-modeled \hi\ (or other Galactic gas), as well as any mis-modeling in its line of sight distance, spin temperature, and spectral index variations. The right panel of Figure~\ref{fig:FM31_dust} shows the results for the baseline fit with the arc template. 

We find that the specialized IEMs for the analysis of FM31, both the baseline model and the baseline model with the arc template, yield an extended excess at the level of $\sim$5\% in the $\sim$3--20 GeV energy range. We show that the excess is also present with similar characteristics when alternative IEMs are employed, and when systematic variations of the spectrum of point sources are considered.

To determine whether the excess presents a spherically symmetric gradient about the center of M31, which would lend support to the hypothesis that it originates from there (at least partially), we perform a further fit in FM31 by including three symmetric uniform templates centered at M31. This also allows us to quantify the spectrum and gradient of the positive residual emission. The templates are fit concurrently with the other components of the baseline IEM, including the arc template. The inner disk (inner galaxy) has a radial extension of 0.4$^\circ$ (5.5 kpc projected radius). This is the best-fit morphology as determined in~\citet{Ackermann:2017nya}, and it corresponds to the bright $\gamma$-ray emission towards M31's inner galaxy. The intermediate ring (spherical halo) has a radial extension from $0.4^\circ < r \leq 8.5^\circ$ (117 kpc projected radius). This extension excludes most of the residual emission associated with the arc template, while also enclosing a majority of M31's globular cluster population~\citep{galleti20042mass,huxor2008globular,peacock2010m31,Mackey:2010ix,veljanoski2014outer,huxor2014outer} and dwarf galaxy population~\citep{McConnachie:2012vd,martin2013pandas,collins2013kinematic}, as well as the M31 cloud~\citep{blitz1999high,kerp2016survey}, which is a highly extended lopsided gas cloud centered in projection on M31, possibly associated with the M31 system. The outer ring (far outer halo) covers the remaining extent of FM31, corresponding to a total projected radius of $\sim$200 kpc, and likewise it begins to approach the MW plane towards the top of the field. Boundaries for these components are overlaid to the residuals in Figure~\ref{fig:spatial_residuals_FM31_tuned}. We find that all templates are significantly detected (with a significance of $\geq 5 \sigma$). Furthermore, the M31-related components are able to flatten the positive residual emission in the fractional energy residuals.

The spectrum and intensity for the inner galaxy are consistent with previously published results. The spherical halo and far outer halo have intensities that are much dimmer than the inner galaxy, and present a mild intensity gradient, tapering off with distance from the center of M31, as shown in the left panel of Figure~\ref{fig:M31_radial_profile}. Their spectra are significantly different from all the other extended components in FM31. They peak between $\sim$5--10 GeV, and drop off below and above these energies more steeply than all other contributions. We find it difficult to reconcile these spectra with the possibility that the excess emission originates solely within the MW, further setting it apart from known Galactic sources. Beyond these general features, the spectra for the two outer annuli differ from each other with the far outer halo presenting a harder spectrum at low energies. The best-fit spectra for the FM31 spherical halo and far outer halo components are shown in the right panel of Figure~\ref{fig:M31_radial_profile}. We compare the spectral shapes to the systematic band of the Galactic center excess (for an arbitrary normalization) from~\citet{TheFermi-LAT:2015kwa}, and find that they are qualitatively consistent, as can be seen in the Figure. 

Our results show that if the excess emission originates (at least partially) from the M31 system, its extension may reach a distance upwards of $\sim$120--200 kpc from the center of M31. This is consistent with the expectation for a DM signal, as the virial radius for the DM halo extends at least this far. To test this interpretation, in the left panel of Figure~\ref{fig:M31_radial_profile} we compare the radial profile of the observed excess with predictions for a DM signal, including both the M31 halo and the MW halo along the line of sight, with a spectrum and annihilation cross-section consistent with a DM interpretation of the GC excess. We consider different assumptions for the amount of DM substructure in M31 (and the MW), and we find that if a cold DM scenario is assumed that includes a large boost factor due to substructures, the observed excess emission is consistent with this interpretation. Granted, however, the exact partitioning of individual contributions to the signal remains unclear, i.e.\ primary emission from M31's DM halo, secondary emission in M31, emission from the local DM filament between M31 and the MW, and emission from the MW's DM halo along the line of sight. This is an intriguing finding, however, its implications are far reaching, and better understanding the MW foreground, as well as complementarity with other DM targets, is crucial before drawing any stronger conclusions. 

\section{Summary}
We present the first search for extended emission from M31 in $\gamma$-rays out to a distance of $\sim$200 kpc from its center. We find evidence for an extended excess that appears to be distinct from the conventional MW foreground, having a total radial extension upwards of 120--200 kpc from the center of M31. We discuss plausible interpretations for the excess emission, but emphasize that uncertainties in the MW foreground, and in particular modeling of the \hi-related components, have not been fully explored and may impact the results. We find that a DM interpretation provides a good description of the observed emission and is consistent with the GC excess DM interpretation. However, better understanding of the systematics, and complementarity with other DM searches, is critical to settle the issue. 

These results were first presented in a poster at the 8th International Fermi Symposium, Oct.~14-19, 2018, Baltimore, MD. The article describing the full analysis is under preparation and will be published elsewhere.

\section*{Acknowledgements}

The authors thank Tsunefumi Mizuno, Gulli J\'ohannesson, Alex Drlica-Wagner, and Troy Porter for many useful comments made at the preparation stage of the manuscript. The authors are also pleased to acknowledge conversations with Ketron Mitchell-Wynne, Sean Fillingham, Tim Tait, Philip Tanedo, Mike Cooper, James Bullock, Manoj Kaplinghat, Kevork N. Abazajian, Sebastian Trojanowski, Ferdinand Badescu, Volodymyr Takhistov, Deano Farinella, and Dan Hooper. A majority of the data analysis has been performed on UCI's HPC, and CK thanks Harry Mangalam for his assistance on numerous occasions. CK also thanks James Chiang for his assistance with the Fermi Science Tools. The work of CK and SM is supported in part by Department of Energy grant DESC0014431. SSC is supported by National Science Foundation Grant PHY-1620638 and a McCue Fellowship. IM acknowledges partial support from NASA grant NNX17AB48G.

The \textit{Fermi}-LAT Collaboration acknowledges support for LAT development, operation and data analysis from NASA and DOE (United States), CEA/Irfu and IN2P3/CNRS (France), ASI and INFN (Italy), MEXT, KEK, and JAXA (Japan), and the K.A.~Wallenberg Foundation, the Swedish Research Council and the National Space Board (Sweden). Science analysis support in the operations phase from INAF (Italy) and CNES (France) is also gratefully acknowledged. This work performed in part under DOE Contract DE-AC02-76SF00515.

\bibliography{citations}
\end{document}